\documentclass[10pt,conference,letterpaper]{IEEEtran}
\IEEEoverridecommandlockouts
\pdfoutput=1
\usepackage[T1]{fontenc}
\usepackage{cite}
\usepackage{amsmath,amssymb,amsfonts}
\usepackage{algorithm}
\usepackage[noend]{algpseudocode}
\usepackage{flushend}
\usepackage{graphicx}
\usepackage{textcomp}
\usepackage{dcolumn} 
\usepackage{bm} 
\usepackage{braket}
\usepackage{hyperref} 
\hypersetup{
    colorlinks = true,
    citecolor = magenta,
    linkcolor = purple
}
\def\BibTeX{{\rm B\kern-.05em{\sc i\kern-.025em b}\kern-.08em
    T\kern-.1667em\lower.7ex\hbox{E}\kern-.125emX}}
    
\usepackage[svgnames, table]{xcolor}
\usepackage{tabularx, makecell, linegoal}

\usepackage{ifthen}

\newboolean{ShowComments}
\setboolean{ShowComments}{true}  
\ifthenelse{\boolean{ShowComments}}%
	{
		\newcommand{\ColorComment}[3]{%
				{\colorbox{#1}{\color{White}   \textsf{\textbf{#2}}} \textcolor{#1}{#3}}}

	}%
	{
		\newcommand{\ColorComment}[3]{}

	}%

\definecolor{rdvcolor}{rgb}{0,0.5,0}
\definecolor{satohcolor}{RGB}{254,0,0}
\definecolor{michalcolor}{RGB}{255,127,80}
\definecolor{naphanncolor}{RGB}{112, 51, 173}

\begin{document}
\bstctlcite{IEEEexample:BSTcontrol}

\title{Architecture and protocols for all-photonic quantum repeaters
\thanks{
This work was supported by JST [Moonshot R\&D] [JPMJMS226C].
}
}

\author{
\IEEEauthorblockN{
Naphan Benchasattabuse\IEEEauthorrefmark{1}\IEEEauthorrefmark{3},
Michal Hajdu\v{s}ek\IEEEauthorrefmark{1}\IEEEauthorrefmark{3},
and Rodney Van Meter\IEEEauthorrefmark{2}\IEEEauthorrefmark{3}}\\

\IEEEauthorblockA{\IEEEauthorrefmark{1}\textit{Graduate School of Media and Governance, Keio University Shonan Fujisawa Campus, Kanagawa, Japan}}
\IEEEauthorblockA{\IEEEauthorrefmark{2}\textit{Faculty of Environment and Information Studies, Keio University Shonan Fujisawa Campus, Kanagawa, Japan}}
\IEEEauthorblockA{\IEEEauthorrefmark{3}\textit{Quantum Computing Center, Keio University, Kanagawa, Japan}\\
\{whit3z,michal,rdv\}@sfc.wide.ad.jp}
}

\thispagestyle{plain}
\pagestyle{plain}
\maketitle

\begin{abstract}
The all-photonic quantum repeater scheme, utilizing a type of graph state called the repeater graph state (RGS), promises resilience to photon losses and operational errors, offering a fast Bell pair generation rate limited only by the RGS creation time (rather than enforced round-trip waits).
While existing research has predominantly focused on RGS generation and secret key sharing rate analysis, there is a need to extend investigations to encompass broader applications, such as distributed computation and teleportation, the main tasks envisioned for the Quantum Internet.
Here we propose a new emitter-photonic qubit building block and a RGS protocol that addresses several key considerations: end node involvement in connection establishment, decoding of logical qubits within the RGS, and computing the Pauli frame corrections at each participating node to ensure the desired correct end-to-end Bell pair state.
Our proposed building block significantly reduces the total number of emissive quantum memories required for end nodes and seamlessly integrates all-photonic and memory-based repeaters under the same communication protocol.
We also present an algorithm for decoding logical measurement results, employing graphical reasoning based on graph state manipulation rules.
\end{abstract}


\section{Introduction}
\label{sec:introduction}

Quantum repeaters are the key components to realize the Quantum Internet~\cite{wehner-vision-road-ahead, rfc9340, michal-rdv-qc-book, azuma-rmp-repeater-review}.
They are used to circumvent photon losses due to attenuation in optical fibers over long distances via splitting the distance the photon needs to travel into multiple segments and later splicing them into one long-distance entangled quantum state shared between distant parties~\cite{briegel-repeater}.
The traditional approach to quantum repeaters requires the use of quantum memories to store the shorter-length entanglement before splicing them into an end-to-end entanglement~\cite{duan-repeater, sangouard-repeater}.
In contrast, a recently proposed quantum repeater scheme based on the properties of graph states, also known as all-photonic quantum repeaters, does not rely on quantum memories~\cite{azuma-rgs, borregaard-3g-repeater, rozpedek-all-photonic-bosonic-gkp-code, niu-all-photonic-one-way-repeater}.
This approach utilizes highly entangled photonic states, increasing the redundancy in the quantum states, and allowing the states to be corrected and turn into the desired entangled states as long as a certain subset of the photons distributed between adjacent repeaters along the end-to-end path arrive.

Although all-photonic quantum repeater schemes promise many attractive features such as the fast generation of Bell pairs (limited only by the state generation) and tolerance of both operational errors and losses, they also come with new challenges. 
One of the main challenges that many groups are tackling is how the highly entangled photonic states can be efficiently generated with regard to various metrics.
This problem is being tackled from both the theoretical~\cite{pant-rate-dist-tradeoff, buterakos-graph-generation, zhan-graph-gen-delay-line,  shapourian-graph-gen, russo-arbitrary-graph-state, hilaire-rgs-optimizing-gen-time, li-entangled-photon-factory, kaur-patil-guha-rgs-generation, ghanbari-hoi-kwong-rgs-optimization} and the experimental perspectives~\cite{hasegawa-passive-absa-experiment, jian-wei-pan-rgs-experiment, lodahl-entangle-photonic-fusion}.
Other open challenges, especially regarding the architecture and the operation of the network, such as how memory-equipped end nodes should participate in the all-photonic schemes, how protocols decide who should act on the information of the measurements to get a deterministic Bell pair, or how to integrate the all-photonic repeaters with memory-based repeaters under the same protocols, have gone largely unexplored~\cite{naphan-rgs-tutorial}.

The core principle of quantum repeaters, entanglement swapping, is essentially quantum teleportation~\cite{bennet-teleportation} performed on one half of a generic Bell pair state.
Thus, to complete the swap, measurement results need to be sent to the two parties sharing the swapped Bell pair to apply correction operations to obtain a previously agreed-upon type of Bell pair.
All-photonic quantum repeaters replace quantum memories with highly entangled photons, some of which will be lost.
Given that the ratio of photons to end-to-end Bell pairs can span orders of magnitude, accurately tracking the state becomes a non-trivial task.

In this work, we build upon the all-photonic scheme proposed by Azuma \emph{et al.}~\cite{azuma-rgs}, which is based on the repeater graph state (RGS), and present two contributions.
First, we introduce a building block called ``half-RGS,'' generated via deterministic quantum emitter method~\cite{buterakos-graph-generation}, which serves as an emitter anchoring photonic qubits in a graph state.
This building block can be integrated into both the RGS source nodes and the end nodes, enabling applications requiring corrected Bell pairs beyond quantum key distribution (QKD).
Utilizing this building block at end nodes also reduces the number of emissive quantum memories required to be held at those nodes, enabling them to keep up with the high trial rate of Bell pair generation provided by the RGS scheme.
It also bridges the disparity between all-photonic and memory-based repeaters to operate under the same network protocol by treating the all-photonic segments as one long memory-based link, making it possible for the first time to connect an RGS-based network to a quantum computer or multicomputer interconnect.

Second, we propose a communication protocol outlining which classical data needs passing between network nodes, as well as the methods for state tracking and calculation of the Pauli frame corrections.
Through simulations of the correction procedure using a stabilizer simulator~\cite{gidney-stim}, we confirmed that our protocol calculates the correct operations at end nodes to get Bell pairs in the desired state.

In Sec.~\ref{sec:graph-states}, we give a brief introduction to graph states, and explain graph manipulation rules, which we will use to reason about the correctness of the protocol.
Equipped with this knowledge, we then describe the all-photonic repeater scheme based on the RGS in Sec.~\ref{sec:rgs-scheme}.
Readers familiar with all-photonic repeaters can skip these sections, but may find them useful (especially for our implementation-focused terminology) as we discuss our contributions beginning in Sec.~\ref{sec:rgs-scheme}. 
We then propose in Sec.~\ref{sec:half-rgs} a new emitter-photon graph state building block, called the half-RGS, that supports the integration of RGS scheme-type links into the memory-equipped end nodes and memory-based repeater networks.
Generation of the RGS from the half-RGS and its generation based on our deterministic quantum emitter approach, including the Clifford side effects that are generated by each vertex in the graph state, is discussed in Sec.~\ref{sec:generation-of-half-rgs}.
Our protocol defining which information should be sent to end nodes, which information can remain local, and how side effects and measurement results are interpreted at the ABSA and then at end nodes to get a corrected Bell pair is porposed in Sec.~\ref{sec:protocol}.
Finally, we discuss the implications and possible extensions of the architecture and the protocol in Sec.~\ref{sec:discussion}.

\section{Graph States}
\label{sec:graph-states}

To understand our proposed protocol and the generation of graph states via the deterministic generation with quantum emitter~\cite{buterakos-graph-generation}, it is best to be familiar with the graph state formalism.
Here, we provide a brief description of graph states and the required manipulation rules.
For a more detailed reading of graph states and their application, we refer readers to~\cite{hein-graph-state-pra, hein-graph-state-arxiv, patil-guha-clifford-manipulation}.

\subsection{Graph State Definition}

Graph states~\cite{briegel-raussendorf-graph-state} are a class of quantum states that can be represented via a simple graph $G(V, E)$ where $V$ is the set of vertices and $E$ is the set of edges connecting vertices in $V$.
In the graph state formalism, a vertex represents a qubit initialized in $\ket{+}$ while an edge represents an interaction between two qubits given by the controlled phase gate $CZ = \ket{0}\bra{0} \otimes I + \ket{1}\bra{1} \otimes Z$. 
Thus a graph state defined from $G(V, E)$ can be described as
\begin{equation}
    \ket{G(V, E)} = \prod_{\{u, v\} \in E} CZ(u, v) \ket{+}^{\otimes |V|}.
\end{equation}

Graph states are a subclass of more general stabilizer states~\cite{gottesman-stabilizer}.
They can be described in terms of stabilizer generators,
\begin{equation}
    g_u = X_u \prod_{v \in N_u} Z_v
\end{equation}
associated with each vertex $u$ in $V$, where $N_u$ is the set of neighbor vertices of $u$.

\begin{figure*}[htb]
    \centering
    \includegraphics[width=\textwidth]{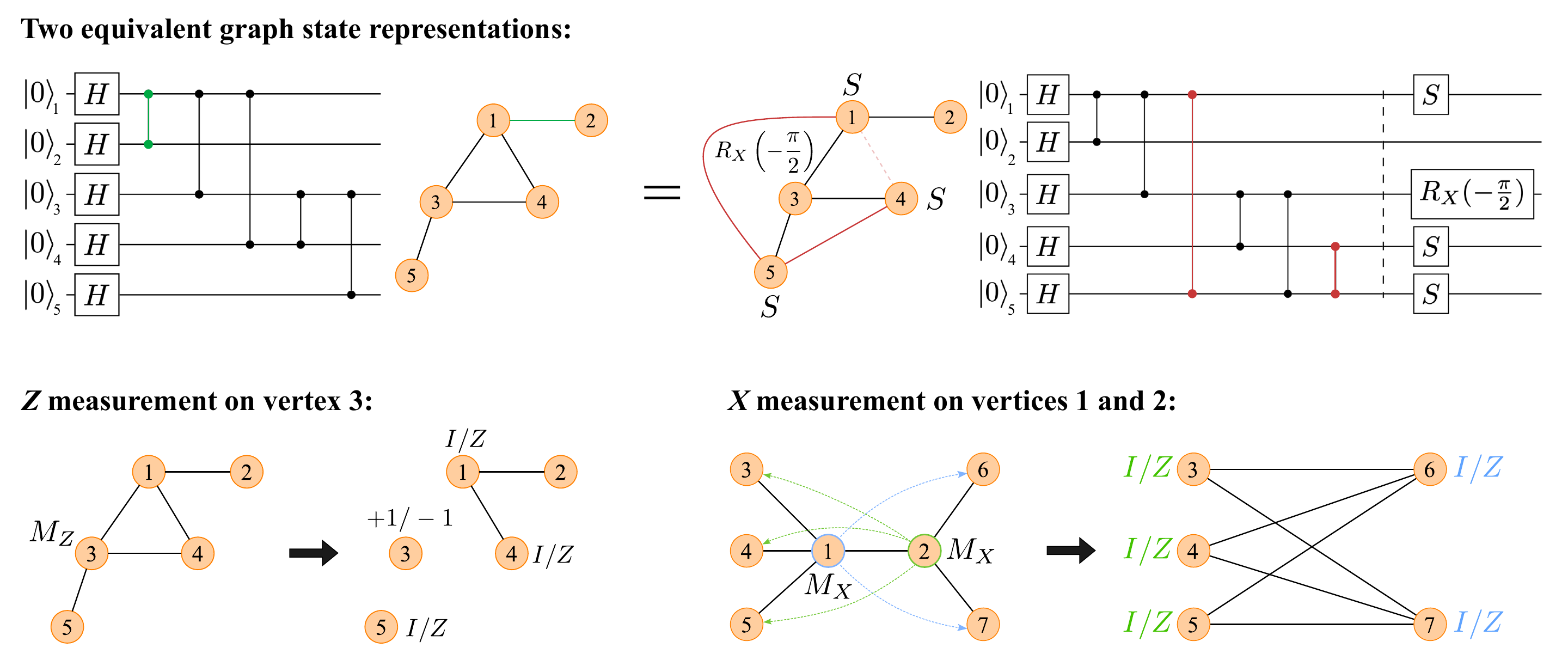}
    \caption{Two graph state representations of the same quantum state are depicted at the top. Here, vertices represent qubits in the quantum circuit, and the application of a controlled-phase gate between two qubits corresponds to an edge between the corresponding vertices. An example of such an edge is highlighted in green in the top left. The graph state on the top right is obtained by altering the description through local complementation on vertex 3. This process deletes any existing edges between neighboring vertices of vertex 3, such as the edge $(1,4)$ in this case, and introduces new edges that were previously absent, highlighted in red as $(1,5)$ and $(4,5)$. The application of the depicted Clifford operations ensures that despite the differing graph representations, the quantum states remain identical in both cases. Visualization of a Z measurement with side effects is presented in the bottom left, while the impact of two X measurements (XX measurement) on a different graph is illustrated in the lower right. The Clifford side effects ($I/Z$) of qubits 3, 4, and 5 depend on the measurement outcome of qubit 2, while those of qubits 6 and 7 hinge on the outcome of qubit 1, as indicated by the blue and green arrows.}
    \label{fig:graph-state-example}
\end{figure*}

It has been shown that all stabilizer states are equivalent to some graph states up to local Clifford operators~\cite{van-den-nest-stabilizer-to-graph, grassl-stabilizer-code-to-graphical-code, schlingemann-stabilizer-code-to-graph-code}.
In other words, for any stabilizer state $\ket{\psi}$, there is a (possibly non-unique) graph state description $\ket{G, \underline{C}}$,
\begin{equation}
    \ket{\psi} = \ket{G; \underline{C}} = \ket{G; C_1, C_2, \ldots, C_n} = \bigotimes_{i=1}^n C_i \ket{G},
\end{equation}
where $C_i \in \mathcal{C}_1$ is a single qubit Clifford gate acting on qubit $i$, with $\mathcal{C}_1$ being the local Clifford group (single qubit Clifford gates).
We call the Clifford operator acting on each vertex of the graph the ``Clifford side effect'' or just ``side effect'' when the context is clear.
These side effects when drawn are represented by a letter next to the vertex as shown in Fig.~\ref{fig:graph-state-example}.

\subsection{Measurements on Graph States}

One appealing property of performing Pauli basis measurements on a qubit within a graph state is that the result is still a graph state with changes to the Clifford side effects of neighbor vertices of the measured qubit in the graph depending on the measurement outcomes.
In this work, we require two sequences of measurements; the Z measurement and the XX measurements (as depicted in Fig.~\ref{fig:graph-state-example}).
These two measurements introduce possible Z side effects to the resulting graph.
Since only Z side effects are added or removed, it is simple to track the side effect of each vertex with just one bit of information.

Measuring a qubit with $Z$ side effect in the Z basis does not change the measurement outcome since the Z gate and the Z measurement commute.
On the other hand, measuring a qubit in the X basis when there is a $Z$ side effect on the qubit flips the result since $P_{X, \pm}(a) Z_a \ket{G} = P_{X, \mp}(a) \ket{G}$, where $P_{X, \pm}(a)$ denotes the projection of qubit $a$ onto the eigenvectors with eigenvalue $\pm 1$ of the Pauli operator $X$.
These properties are crucial to note since they affect how the side effects are propagated during the entanglement distribution in the RGS scheme.

\section{all-photonic quantum repeater}
\label{sec:rgs-scheme}

In this section, we review the all-photonic scheme proposed by Azuma \emph{et al.}~\cite{azuma-rgs}.
We will refer to this scheme as ``RGS scheme'' from here onwards.

\subsection{RGS Scheme Overview}

\begin{figure*}[htb]
    \centering
    \includegraphics[width=\textwidth,keepaspectratio]{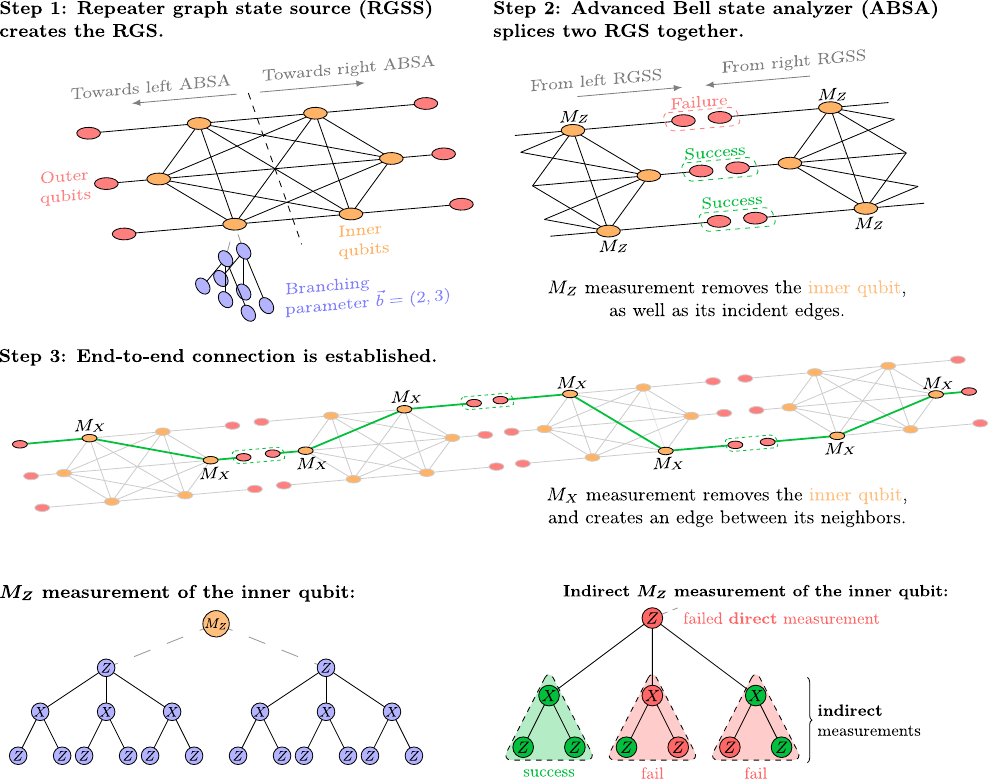}
    \caption{An overview of the RGS scheme. The three steps shown here have corresponding actions to the memory-based repeater scheme, where inner encoded qubits correspond to the memories while outer qubits correspond to the emitted photons. RGS generation in step 1 (at RGSS) mirrors the entanglement swapping of quantum memories but without actually choosing which inner qubits will be paired up. Step 2 (at ABSA) illustrates the link-level generation process through the BSM between each pair of outer qubits. The Z measurement on inner qubits in step 2 and the X measurements in step 3 signify the choosing of which pairs are swapped. Logical measurement of inner qubits in the Z basis is depicted at the bottom. The $Z$ and $X$ labels inside the blue physical qubits indicate the actual physical measurement bases. For logical X measurement, the Z and X measurement bases of physical qubits are swapped. The indirect Z basis measurement of a physical qubit in the tree encoding is shown in the bottom right. If a direct Z measurement on a qubit fails due to photon loss, the result can still be inferred from the eigenvalue parity of qubits within any of the dotted triangles.}
    \label{fig:rgs-scheme-overview}
\end{figure*}

The RGS is made of $2m$ \emph{inner qubits}, forming a complete graph, and $2m$ \emph{outer qubits}, linked to an inner qubit, as depicted in Fig.~\ref{fig:rgs-scheme-overview}.
This complete graph of inner qubits encodes the entanglement swap, and is done before the RGS are sent out to the measurement devices generating the link-level entanglement, thus it is considered a ``time-reversed'' procedure.

In the RGS scheme, outlined in Fig.~\ref{fig:rgs-scheme-overview}, there are two node types. 
The RGS source nodes (RGSS) and the advanced Bell state analyzer nodes (ABSA) function similarly to the typical memory-equipped repeaters and the BSA nodes, respectively.
To create an end-to-end Bell pair, all the RGSSs between the two end nodes create the RGS.
Each of the RGS is split in half, and each half is then sent to its adjacent ABSA.
Bell state measurements (BSM) are performed on every pair of the outer qubits between the two halves of the RGS meeting at each ABSA.
A successful BSM between the outer qubits creates an entanglement link between their connected inner qubits.
Aside from the two inner qubits that are now connected, single qubit Z basis measurements are then performed on the rest of the inner qubits, removing them from the system, and leaving a linear chain graph state between the two end nodes.
Finally, XX measurement is performed on the remaining two inner qubits at each ABSA, creating an end-to-end two-vertex graph state (a rotated Bell pair).

\subsection{Loss Tolerance of the RGS Scheme}

The RGS effectively combats the photon loss by increasing the number of arms (the parameter $m$), which in turn improves the probability of obtaining at least one successful BSM and the probability of getting an end-to-end Bell pair.
However, blindly increasing the number of arms is not the best approach, as successful measurements of all the inner qubits are needed.
Thus the inner qubits can be encoded in the tree graph state encoding~\cite{varnava-counterfactual-measurement}, a form of quantum error-correcting code, and logical measurements of inner qubits can be done via a process called \emph{counterfactual measurement} or \emph{indirect measurement}~\cite{varnava-counterfactual-measurement}.
The number of arms $m$, and the branching parameter $\vec{b} = (b_0, b_1, \ldots, b_{n-1})$, which describes the tree encoding, together determine how much loss and error the RGS can tolerate.
An example of the tree code encoded inner qubits with $\vec{b} = (2, 3)$ is shown in Fig.~\ref{fig:rgs-scheme-overview}.

\subsection{Logical Measurement of Inner Qubits}

The logical measurement of inner qubits can be achieved by performing single qubit measurements on all of the physical qubits in a pre-decided pattern, where the qubits in the same level of the tree undergo the same basis of measurements (depicted in Fig.~\ref{fig:rgs-scheme-overview}).
For the X (Z) basis measurement, perform the X (Z) measurement on the first level of the tree and alternate into the Z (X) measurement on the second level.
The alternation of the basis used on the tree level is done for every level, meaning that for the X (Z) basis measurement, the X (Z) measurement is done on the odd level while the Z (X) measurement is done on the even level.

The concept behind the counterfactual measurement stems from the fact that graph states can be described using the stabilizer formalism.
The measurement result of qubit $a$, which is part of the support of the stabilizer generator $g_i$, can be deduced if all other qubits in the generator $g_i$ were successfully measured (either directly or indirectly).
Therefore, the Z measurement result of a qubit $a$ at level $k$ can be deduced if \emph{any} of its neighbors $b$ at level $k+1$ are successfully directly measured in the X basis and \emph{all} of $N_b$ in level $k+2$ are also successfully measured in the Z basis (either directly or indirectly).

The logical measurement results of the inner qubits can be inferred from finding the eigenvalue parity of a specific set of obtained physical results.
For Z basis measurements, the logical outcome is determined by the parity of eigenvalues across all qubits in the first level. Conversely, for X basis measurements, the logical outcome is determined from the parity of the eigenvalue resulting from the X measurement of qubit $a$ in the first level and the eigenvalues of its neighbors measured in the Z basis in the second level (either direct or indirect). 
In scenarios involving only photon losses, all the logical results deduced from any qubit in the first level and their neighbors are all identical. 
For proof of how the parity of these groups of qubits signifies the logical result, we refer readers to the supplementary material of~\cite{azuma-rgs}.

\subsection{Operational Error Tolerance of RGS}

The tree encoding also offers tolerance to quantum operation errors.
To correct physical measurement results of qubits in the tree, only those measured in the Z basis can be corrected.
Correcting a qubit at level $k$ can be achieved by performing a majority vote on the measurement result of the neighbor qubits at level $k+1$ and its neighbors at level $k+2$. 
Meanwhile for the logical results, only the inner qubits measured in the X basis can be corrected.
This is done by performing a majority vote on the X measurement result of all the qubits in the first level since only one of them suffice for deducing the logical result.
This is in contrast to the logical Z, which necessitates success across all Z measurements (either direct or indirect) in the first level.
These properties imply that the capacity to tolerate errors of RGS also depends on the loss probability of photons.

\subsection{Transmission Order of Photons in the RGS}

The measurement basis selection of the inner qubits is dependent on the BSM outcome of the outer qubits.
As long as the outer qubits arrive at the ABSA before their connected inner qubit, the measurement basis can be determined locally by the ABSA.
The physical qubits composing the inner qubits can be sent in any order as long as it is known to the ABSA beforehand, this is fixed and does not require any delay line to correct the ordering in our work as we will show the generation sequence in Sec.~\ref{sec:generation-of-half-rgs}.
We note that even if the photons are lost, assuming that the photons are well separated temporally, the ABSA can deterministically flag the loss event.
This well-separated assumption is also commonly adopted in memory-based repeater schemes for multiplexing photons from multiple memories into a single fiber~\cite{rdv-qi-architecture}.

\section{Half-RGS Building Block}
\label{sec:half-rgs}

\begin{figure}
    \centering
    \includegraphics[width=0.8\columnwidth]{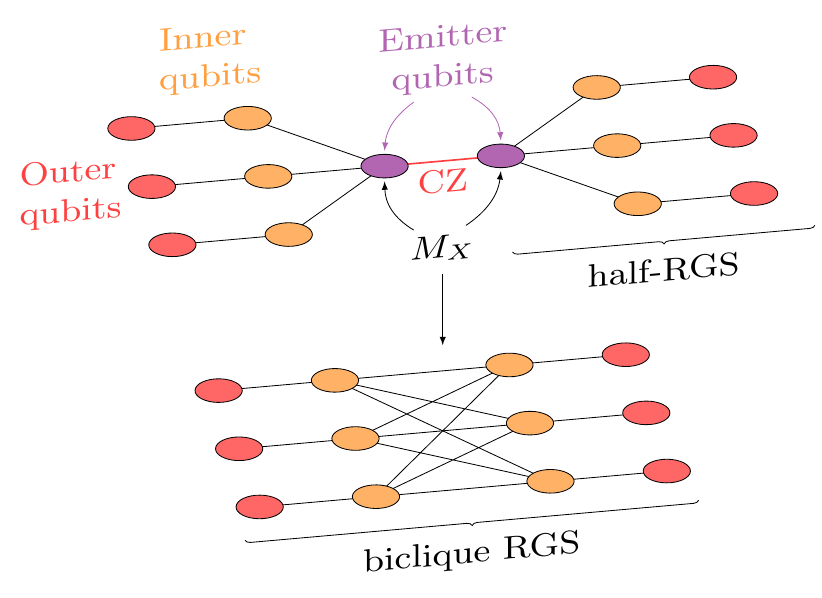}
    \caption{An example of half-RGS and the transformation of two half-RGSs into a biclique RGS. The anchor emitter qubits of the two half-RGSs are joined via an application of a CZ gate follwed by the XX measurement resulting in a biclique RGS.}
    \label{fig:half-rgs-biclique-rgs}
\end{figure}

The complete graph connectivity of inner qubits in the RGS is not a strict requirement to encode the entanglement swap.
If each half of the inner qubits is connected into a complete bipartite graph (biclique) as depicted in Fig.~\ref{fig:half-rgs-biclique-rgs}, this is enough to be used as the RGS~\cite{russo-rgs-biclique-generation, tzitrin-rgs-biclique-equivalent}.
From here on, we will refer to this biclique RGS also as the RGS when it is unnecessary to differentiate between them.

In this section, we present our first contribution, the \emph{half-RGS}, which we use as our building block to create the biclique RGS.
It bridges the end nodes and RGSS disparity and also allows seamless integration between memory-based repeaters and RGS-based repeaters.

\subsection{Half-RGS Description}

The \emph{half-RGS} is depicted in Fig.~\ref{fig:half-rgs-biclique-rgs}. 
The blue qubit denotes an emitter (or a memory qubit) acting as an anchor to half of the photonic RGS.
Joining two of them at the anchors via a CZ gate and performing the XX measurement result in the RGS, thus the name half-RGS.
Next, we show that some of the previously open problems for the RGS scheme~\cite{naphan-rgs-tutorial} can be addressed by utilizing the half-RGS building blocks.

\subsection{Improving Resource Requirement at End Nodes}

\begin{figure*}
    \centering
    \includegraphics[width=\textwidth]{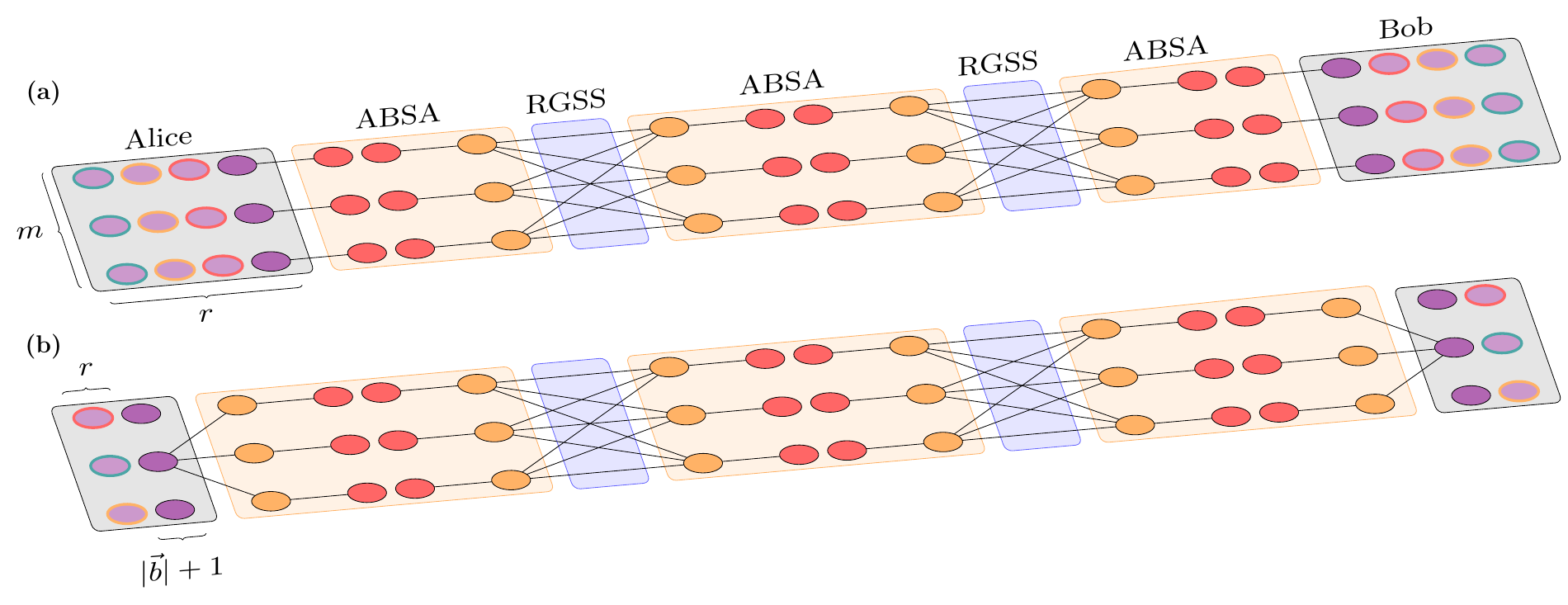}
    \caption{Architectures supporting the RGS scheme where the photonic states generated at end nodes are different. (a) The architecture proposed in~\cite{zhan-graph-based-repeater-analysis}, where end nodes require $m$ emissive memories for each trial, along with $rm$ idle memories that have already been utilized in prior trials and are awaiting messages from all ABSAs. (b) Proposed architecture, featuring end nodes equipped with half-RGS building blocks. In this setup, end nodes require $|\vec{b}|+1$ quantum emitters (where $|\vec{b}|$ represents the depth of the tree encoding plus one), while reserving $r$ memories awaiting notification messages from prior trials. Purple circles with black borders represent the quantum emitters utilized in the current trial. Purple circles with colored borders in both (a) and (b) indicate idle memories awaiting messages from ABSAs, with the same colors denoting memories participating in the same trials.}
    \label{fig:proposed-architecture}
\end{figure*}

The incorporation of end nodes into the RGS scheme has not been explored aside from QKD applications where end nodes are measurement nodes and not full compute nodes.
The main difference in having compute nodes as opposed to measurement nodes is the requirement that compute nodes need to have quantum memories storing the states and wait for all the measurement result messages from ABSA and RGSS to arrive before the Bell pairs are ready for use. 
One way to realize this was explored~\cite{zhan-graph-based-repeater-analysis} (depicted at the top of Fig.~\ref{fig:proposed-architecture}), where the end nodes are equipped with at least $m$ (the number of RGS arms) quantum memories, and for one trial, having $m$ memories emit photons to meet with half-RGS at an ABSA.

In the approach outlined in~\cite{zhan-graph-based-repeater-analysis}, achieving correctable Bell pairs while fully utilizing the RGS scheme's rapid trial time necessitates end nodes to have an additional $rm$ memories in reserve, with photons already emitted from all the memories waiting for the correction messages to arrive.
Here, $r$ denotes the ratio of the time required for the classical message to travel from the farthest ABSA and the RGS generation time.
A rough calculation suggests that for a separation distance of 1000 km, end nodes would require approximately 150 memories, based on the analysis presented in~\cite{hilaire-rgs-optimizing-gen-time}, to maintain pace with the trial rate ($r = 10$, $m = 15$).

In contrast, our proposed architecture (shown at the bottom of Fig.~\ref{fig:proposed-architecture}), leveraging the half-RGS building block, streamlines this requirement by necessitating only $r$ quantum memories and $|\vec{b}| + 1$ emitters (where $|\vec{b}|$ denotes the depth of the tree encoding
at each end node, resulting in a total of $r+|\vec{b}|+1$ emitters.
The $r$ quantum memories, unlike emitters, do not need to emit photons, provided that the qubit state of the anchor emitter of the half-RGS is swapped into these quantum memories. 
This configuration effectively doubles the trial rate compared to the $rm$ memories (or $2rm$ when employing multiplexed fiber to match our achievable rate) approach in~\cite{zhan-graph-based-repeater-analysis}.
The difference in the architecture is illustrated in Fig.~\ref{fig:proposed-architecture}.

\subsection{Integration with Memory-based Repeaters}

\begin{figure*}[htb]
    \centering
    \includegraphics[width=\textwidth,keepaspectratio]{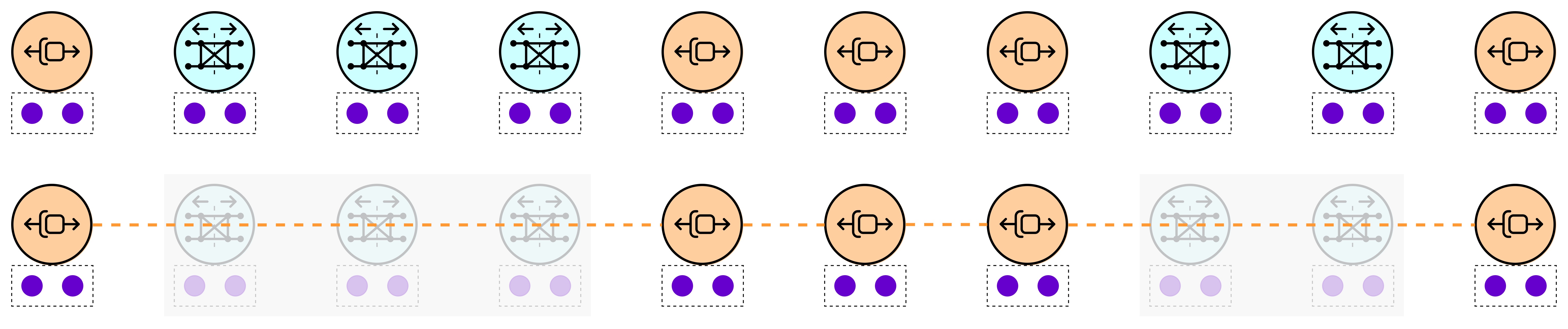}
    \caption{Segments constituting RGSS (repeater graph state source, in light blue) and ABSA (advanced Bell state analyzer, omitted) nodes are treated as virtual links to memory-equipped repeaters (orange). The RGS scheme link is invisible at the connection level protocol, reducing the cost of connection setup and management.}
    \label{fig:rgs-virtual-link}
\end{figure*}

In light of our current discussion, it is more appropriate to consider the RGS scheme as a link-level connection, as depicted in Fig.~\ref{fig:rgs-virtual-link}.
In particular, the RGS scheme can be used to connect any two nodes, not necessarily only end nodes.
The Bell pairs created from the RGS scheme can be treated the same way as link-level resources created from other link architectures, such as Memory-Interference-Memory, Memory-Memory~\cite{sangouard-repeater}, Memory-Source-Memory~\cite{cody-msm}, or Sneakernet links~\cite{simon-sneakernet}, and later can be allocated to the resource management software of the quantum nodes.
This is in the spirit of heterogeneity and interoperability of the future Quantum Internet~\cite{rdv-qi-architecture}.
By abstracting the RGS scheme repeaters into a virtual link-level connection between two memory-equipped nodes, the network can seamlessly operate within the framework of event-driven network protocols proposed in~\cite{cocori-quisp, rdv-qi-architecture, kozlowski-quip-p4}, leading to far greater flexibility.

\subsection{Timing Synchronization}

The original RGS scheme requires that the outer qubits from both sides arrive at the ABSA nearly at the same time, within an indistinguishable time window, for the entanglement swapping via BSM to work.
Since we separate the two halves of RGS to be generated by a different set of emitters, the timing message only needs to be exchanged between neighbors and the controller which controls the emitter at each link in the same way as for conventional memory-based repeaters.
This is unlike the case where the whole graph is generated by one set of emitters~\cite{buterakos-graph-generation, hilaire-rgs-optimizing-gen-time}.
In the single-set case, the timing message must be sent to both left and right nodes to achieve the required coordination along the entire connection path, a significant complication in engineering~\cite{mori2024scalable} that our building block simplifies.

Although this necessitates end nodes to be equipped with a half-RGS generation device, as it is the same component as the RGSS, it is not unreasonable to require it.
A more end nodes friendly approach would be that end nodes connected to a repeater bridging the RGS scheme and the port connecting to end nodes can be memory-based.

\section{Generation of half-RGS}
\label{sec:generation-of-half-rgs}

The generation of RGS is often the most difficult part in realizing the RGS scheme. 
Various generation methods have been proposed~\cite{azuma-rgs, pant-rate-dist-tradeoff, buterakos-graph-generation, zhan-graph-gen-delay-line, shapourian-graph-gen} and optimization techniques building on top to reduce resources and time~\cite{li-entangled-photon-factory, kaur-patil-guha-rgs-generation, ghanbari-hoi-kwong-rgs-optimization}.
Here, we take the approach proposed in~\cite{buterakos-graph-generation} for deterministically generating photonic graph states via quantum emitters.

\subsection{Assumptions on the Operations}

\begin{figure*}[htb]
    \centering
    \includegraphics[width=\textwidth,keepaspectratio]{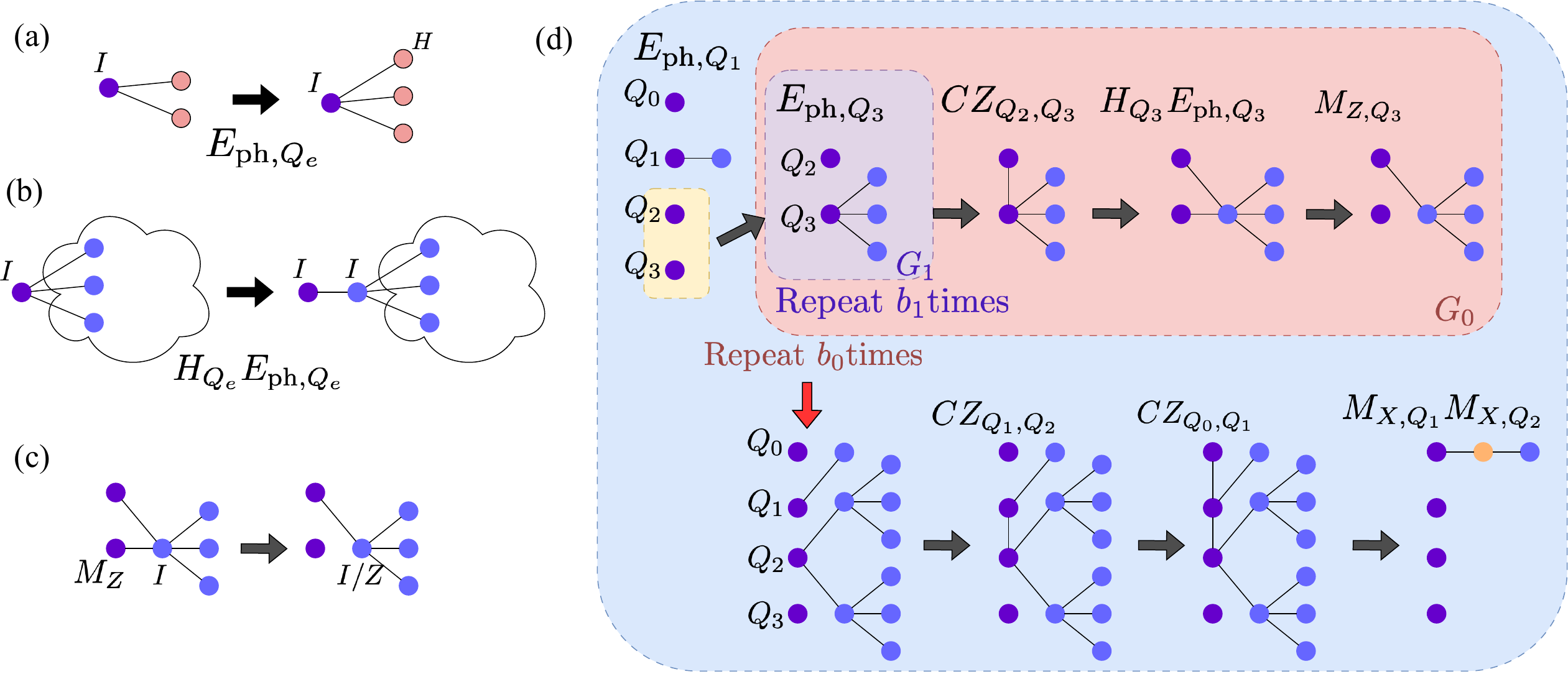}
    \caption{(a) Emission of a photon with side effect labelled. (b) The push-out operation, photon emission followed by a Hadamard gate. This operation does not introduce any side effect. (c) The side effect of push-out followed by a Z measurement. This the main sequence used in generating the half-RGS. (d) The sequence (modified from~\cite{hilaire-rgs-optimizing-gen-time}) to generate one arm of half-RGS with $\vec{b} = (2, 3)$. This needs to be repeated $m$ times to create $m$ arms.}
    \label{fig:rgs-gen-sequence}
\end{figure*}

Assumptions about the quantum emitters and their operations are as follows (also shown in Fig.~\ref{fig:rgs-gen-sequence}).
Quantum emitters are arranged in a linear topology $(Q_0, Q_1, Q_2, \ldots, Q_{|\vec{b}|+1})$.
Hadamard gates can be applied on individual emitters, as well as controlled-phase gates between neighboring ones.
Photon emission is modeled using a controlled-not gate to a new qubit initiliazed in $\ket{0}$ (Fig.~\ref{fig:rgs-gen-sequence}(a)).
In addition to the manipulation rules of graph states mentioned previously, we also require \emph{push-out operations}, where the emitter in the graph state is replaced with the newly emitted photon, and the emitter is now connected to only the new photon~\cite{enocomou-2d-cluster-generation,russo-arbitrary-graph-state}.
This is achieved by photon emission followed by a Hadamard gate, as depicted in Fig.~\ref{fig:rgs-gen-sequence}(b).

The linear optical BSM used in the protocol has the same action as first applying the controlled-phase gate to the two qubits followed by the XX measurement, and has the ability to distinguish only when the eigenvalue product parity is -1, two possible outcomes instead of the four possible outcomes, the theoretical success rate of 50\%.
This choice is arbitrary and any two of the four outcomes would work sufficiently.
An optical single qubit gate can be applied on emitted photons before they leave the RGSS.

\subsection{Sequences for Generating the Half-RGS}

The sequence of generating half-RGS with $m$ arms and branching parameter $\vec{b} = (b_0, b_1, \ldots, b_{n-1})$ anchored at $Q_0$ on one side, a modified version of~\cite{buterakos-graph-generation, hilaire-rgs-optimizing-gen-time}, is given by (also depicted in Fig~\ref{fig:rgs-gen-sequence}(d))
\begin{equation}
  \begin{aligned}
    & \left( M_{X,Q_1} M_{X,Q_2} CZ_{Q_0, Q_1} CZ_{Q_1, Q_2} G_{0} E_{\mathrm{ph},Q_1} \right)^m \\
    & \text { with } G_k = M_{Z, Q_{k+2}} H_{Q_{k+2}} E_{\mathrm{ph}, Q_{k+2}} CZ_{Q_{k+1}, Q_{k+2}} G_{k+1}^{b_{k}} \\
    & \text { and } G_{n-1}=E_{\mathrm{ph}, Q_{n+1}},
  \end{aligned}
\end{equation}
where $E_{\mathrm{ph},Q_i}$ and $M_{B,Q_i}$ denotes photon emission and measurement in the $B$ basis on emitter $i$ respectively. 
To join two half-RGSs into an RGS can be done by performing $CZ$ between $Q_0$ on the left and right side and perform the XX measurement. 

In contrast to the generation sequences and analysis presented in~\cite{hilaire-rgs-optimizing-gen-time}, our approach requires twice the number of emitters per RGSS and an additional two emitters dedicated to the generation of outer qubits.
However, our method achieves a reduction in generation time, halving the previous duration, and eliminates the need for delay lines or efficient photon storage at ABSA.
This lowers the probability of outer qubits being lost in the fiber, as \emph{all} photons are generated in the order of measurements to be performed.

\subsection{Side Effects Created during the Generation}

Both the generation of half-RGS via the quantum emitters at the RGSS and the measurements done at ABSA can introduce Clifford side effects to the state.
Tracking and correcting these side effects are crucial to ensure that the Bell pairs formed between the two end users of the connection are in the desired state.

From the half-RGS generation sequences, shown in~Fig.~\ref{fig:rgs-gen-sequence}, the physical qubits at the lowest level (leaf nodes) of the inner qubits and the outer qubits will have an $H$ side effect (Fig.~\ref{fig:rgs-gen-sequence}(a)).
However, these side effects can be removed via optical Hadamard gates applied to the photons before they are sent out to the ABSA.
On the other hand, the remaining physical qubits in the tree may have a $Z$ side effect introduced with a probability of $50\%$ each, as they are generated via the push-out operation followed by the measurement of its neighbor emitter (Fig.~\ref{fig:rgs-gen-sequence}(c)). 

Furthermore, the outer qubits are also subject to possible $Z$ side effects resulting from the XX measurement connecting them to their respective inner qubits. 

Both the joining of outer and inner qubits and joining two half-RGSs into a biclique RGS involve performing a controlled-phase gate on the two emitters, depicted in the last step of Fig.~\ref{fig:rgs-gen-sequence}(d) and Fig.~\ref{fig:half-rgs-biclique-rgs}, respectively. 
These XX measurements may also toggle the $Z$ side effects on all the physical qubits in the first level of the tree (where the effect was shown in the bottom right of Fig.~\ref{fig:graph-state-example}).
Two approaches are possible to track this; tracking the logical $Z$ side effect on the inner qubit itself, or toggling the side effects of physical qubits between $Z$ and $I$. 
We adopt the latter view as it simplifies the subsequent reasoning, as elaborated in the following sections.

\section{The protocol based on the half-RGS}
\label{sec:protocol}

We now describe the communication protocol to realize the RGS scheme given our assumptions dsicussed above.

\subsection{Overview}

For each trial, every RGSS generates (using the sequence in Fig.~\ref{fig:rgs-gen-sequence}) and sends the RGS to their adjacent ABSA, followed by a classical message containing all the side effects of each physical qubit.
At each ABSA, BSM is performed on the outer qubits, and measurements on the inner qubits are performed, determined by the BSM outcomes.
For each arm of the RGS, ABSA creates a tree of measurement results, depicted in Fig.~\ref{fig:side-effect-propagations-and-corrections}, recording the results of the measurements and the measurement basis, or whether the photon was lost.
The logical measurement results of each inner qubit can then be locally determined at the ABSA by processing all the measurement trees and the side effect information received.
The propagation of the $Z$ side effect to neighbor nodes and eventually to end nodes can be reduced down to two bits denoting whether the logical X result of the inner qubits of the neighbor ABSA should be flipped or not.
One more bit of information is also sent to each end node, denoting the success or failure of trial at each ABSA.
Finally, the end nodes determine the correction operation required to correct the memory held into an end-to-end two-vertex graph state that is equivalent to a Bell pair.

\subsection{Pauli Frame Calculations}
\label{sec:protocol:algorithm}

\begin{figure*}
    \centering
    \includegraphics[width=\textwidth]{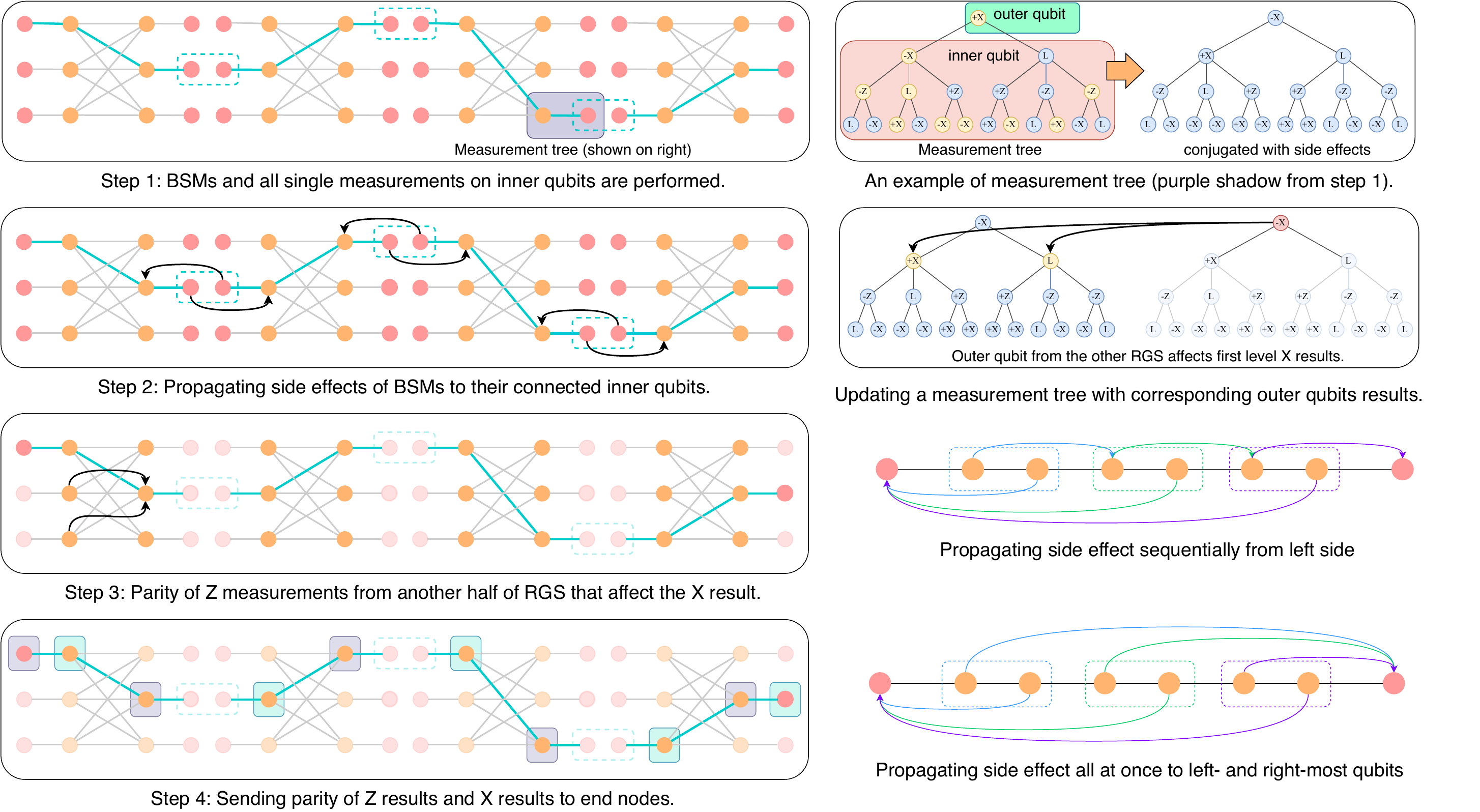}
    \caption{The tracking of side effect propagation at each step, the resolution of physical and logical measurement results, and the rationale behind the Pauli frame corrections applied at end nodes, as illustrated in Steps 1-4 on the left side of the figure. Examples of measurement trees and their updating during the Pauli frame calculation process are presented in the top two panels on the right side. Yellow filled vertex represent that the vertex has a $Z$ side effect. The sequential propagation of side effects, viewed hop-by-hop, is contrasted with an equivalent view of simultaneous propagation to the outermost qubits, depicted in the bottom two panels on the right side.
    }
    \label{fig:side-effect-propagations-and-corrections}
\end{figure*}

Although the photons are measured in the order that they are generated, conceptually, we can think of them being measured in a any order as long as each outer qubit is measured before its corresponding inner qubits.
We now describe how the Pauli frame corrections can be calculated for the memories at the end nodes.
Fig.~\ref{fig:side-effect-propagations-and-corrections} outlines the procedure below.

First, we assume that success or failure of all BSM on the outer qubits are known, thus all the measurement basis on inner qubits have been decided.
Before we begin tracking the state changes due to measurements, ABSAs construct a measurement tree for each inner and outer qubit pair of each RGS arm with the side effect information about all the physical qubits received from RGSS and end nodes.
The measurement tree stores the measurement basis and its result, or marks that the qubit is lost. This is denoted by $\pm X$ or $\pm Z$ for successful measurements, and $L$ for loss, respectively.
If the qubit has a $Z$ side effect from the generation, shown as yellow filled vertex in~Fig.~\ref{fig:side-effect-propagations-and-corrections}, the measurement result in the X basis will be flipped or it can be thought of as conjugating the measurement result with the $Z$ side effect, while lost event result $L$ stays the same.

Next, we propagate the BSM results obtained from the outer qubits to their inner neighbors, as shown in Step 2 of Fig.~\ref{fig:side-effect-propagations-and-corrections}.
BSM has the same effect as an XX measurement in Fig.~\ref{fig:graph-state-example}, and can create a $Z$ side effect.
Successful BSM outcomes propagate to the inner qubits across the RGSs, as shown by the black arrows in Step 2.
If the measurement outcome on the outer qubit from the left (right) side is $-1$ ($-X$ in the figure), the measurement results of the physical qubits in the first level of the right (left) tree are flipped.
We do not need to consider the $Z$ propagation to inner qubits if their outer neighbors were pat of a failed or discarded BSM.

Results of the Z measurements on the inner qubits of the same RGS affect the X measurement results of that RGS.
However, such measurements are performed at different ABSA, making their outcomes locally unavailable.
This is shown in Step 3 in Fig.~\ref{fig:side-effect-propagations-and-corrections}
Fortunately, we can see from the bottom two panels on the right side of Fig.~\ref{fig:side-effect-propagations-and-corrections} that the $Z$ propagation of XX measurements can be processed once at end nodes via the product of the measurement results.
This implies that the Z side effect, possibly occurring from all the inner qubits that have undergone Z measurements, can be dealt with in the same way. ABSAs compute partial parity of locally performed Z measurements, given by the product of the measurement outcomes, and send this information to the end nodes, where it is processed to yield the total parity.

Therefore, each ABSA needs to send only 2 bits of information to the end nodes; 1 denoting whether the $Z$ is propagating to that side of end nodes or not (parity of measurement results), and whether the procedure is successful.
The amount of classical bits that each ABSA receives is the same as the number of total photons.
This is in agreement with the Two-Stage protocol analysis made in~\cite{naphan-rgs-tutorial}.

\subsection{Correctness of the Protocol and the Propagation Rules}

We validated the correctness of the protocol using the stabilizer tableau simulator, Stim~\cite{gidney-stim}.
In the simulation
\footnote{\url{https://github.com/Naphann/repeater-graph-state-protocol-based-on-half-RGS/tree/main}},
each inner qubit does not follow the generation sequence depicted in the pink shadow of Fig.~\ref{fig:rgs-gen-sequence}. 
Instead, they are directly generated from quantum circuits using Hadamard and CZ gates without measurements.
To simulate the stochastic $Z$ side effect at each physical qubit in the inner logical qubit, a $Z$ gate is added with a 50\% probability after the state is generated. Subsequently, each inner-outer qubit arm is generated as described in Sec.~\ref{sec:generation-of-half-rgs} until half-RGSs are obtained, which are then transformed into biclique RGSs.
Once all the RGSs are generated at each RGSS and the two half-RGSs from end nodes, BSMs are performed on all outer qubit pairs, and measurement bases are selected for all inner qubits.
The measurement of all physical qubits in the system is then performed simultaneously (in the same circuit layer).

Instead of directly simulating message passing, the correction procedure is executed step by step, as outlined in Sec.~\ref{sec:protocol:algorithm}. For each RGS arm, a measurement tree is created, followed by updating the measurement tree with side effect information and $Z$ propagation from the BSM, and the decoding of the logical result.
Finally, the 4 bits of information at each ABSA are processed, and correction operations are applied to the end nodes' memories accordingly.

To confirm that the final state between two end nodes is a two-vertex graph state without any side effect, we used Stim to examine the stabilizer generators of the final two qubits.
We verified if the generators are $X_a Z_b$ and $Z_a X_b$ as expected, where the subscripts $a$ and $b$ denote the qubit held by Alice and Bob, respectively.
Simulating the photon loss event is achieved by probabilistically marking qubits as lost with some probability.
If a qubit is marked as lost, we randomly apply Pauli $X$, $Y$, or $Z$ before the measurement, and the results are then removed from the measurement tree, excluding their participation from the logical qubit decoding process.

\section{discussion}
\label{sec:discussion}

In this paper, we proposed a practical communication protocol for the implementation of the all-photonic RGS scheme introduced by Azuma \emph{et al.}~\cite{azuma-rgs}, alongside an architectural framework leveraging the half-RGS building block to generate the photonic graph states.
Previously, all-photonic repeaters have primarily been investigated in the context of QKD applications. 
However, our work extends this paradigm to encompass scenarios necessitating a corrected Bell pair as the ultimate resource, achieved by integrating memory-equipped end nodes into both the protocol and architecture.
This integration allows seamless incorporation of our approach into conventional memory-based repeater networks, consistent with the architectural principle of a quantum internet in~\cite{rdv-qi-architecture}, treating segments between memory-equipped nodes connected via RGS scheme repeaters as virtual physical links at the same abstraction with existing memory-based link architectures.
While our approach adds computational tasks at the ABSA, rather than offloading everything for processing at the end nodes, this is reasonable given the ABSA's existing requirement for fast logic to adaptively select measurement bases. 
Our approach represents a stride toward practical implementation, while also reducing the quantum memory burden at end nodes, thereby fully utilizing the rapid trial rate characteristic of the RGS scheme.

Although the proposed half-RGS possesses a number of desirable characteristics, it remains unclear how it compares with other alternative building blocks. More efficient generation sequences leveraging the local complementation property of graph states could offer advantages such as the inclusion of purification protocols or the extraction of multiple Bell pairs from a single trial.
We note that more efficient generation sequences of the RGS via the quantum emitter have recently been proposed~\cite{kaur-patil-guha-rgs-generation, ghanbari-hoi-kwong-rgs-optimization} during the preparation of this paper with different architecture designs and possibly different Pauli frame correction calculations.
A detailed analysis of our scheme and a comparison with these alternative methods are deferred to future work.

While studies like~\cite{azuma-rgs, hilaire-rgs-optimizing-gen-time, zhan-graph-based-repeater-analysis} have investigated the success probability and error analysis of the RGS scheme under conditions of uniform RGSS and ABSA placement, and uniform noise characteristics, the robustness of the RGS scheme under complex noise models and real-world constraints remains unclear.
Challenges such as diverse link characteristics, uneven node placement, and complex traffic models are not amenable to analytic studies and require simulation.
However, existing quantum network simulators~\cite{cocori-quisp, wu2021sequence, coopmans2021netsquid} do not support the simulation of quantum states comprising hundreds or thousands of entangled qubits. Overcoming this limitation to enable the study of the RGS scheme on quantum network simulators will require innovative approaches.
Lastly, a detailed analysis of generation time and a performance evaluation comparing mixed RGS and memory-based architectures against RGS-only and memory-only configurations remains a pressing open question.

\bibliographystyle{IEEEtran.bst}
\bibliography{IEEEabrv, bibfile}

\end{document}